\documentclass[aps,floatfix,showpacs,twocolumn,superscriptaddress]{revtex4}  
\usepackage{graphicx} 
\usepackage{dcolumn} 
\usepackage{amsmath} 
\usepackage{color}

\begin{document}
\newcommand{\bce}{\begin{center}} \newcommand{\ece}{\end{center}}
\newcommand{\beq}{\begin{equation}} \newcommand{\eeq}{\end{equation}}
\newcommand{\beqy}{\begin{eqnarray}}
\newcommand{\eeqy}{\end{eqnarray}}
\def\be{\begin{eqnarray}}
\def\ee{\end{eqnarray}}
\def\bp{{\mbox{\boldmath $p$}}}
\def\bP{{\mbox{\boldmath $P$}}}
\def\bK{{\mbox{\boldmath $K$}}}
\def\br{{\mbox{\boldmath $r$}}}
\def\bq{{\mbox{\boldmath $q$}}}
\def\bR{{\mbox{\boldmath $R$}}}
\def\qvec{{\mbox{\boldmath $q$}}}
\def\kvec{{\mbox{\boldmath $k$}}}
\def\nablavec{{\mbox{\boldmath $\nabla$}}}
\def\sigmavec{{\mbox{\boldmath $\sigma$}}}
\def\Deltavec{{\mbox{\boldmath $\Delta$}}}
\newcommand{\no}{\noindent}
\newcommand{\Tr}{\rm Tr}
\newcommand{\anu}{\bar\nu}
\newcommand{\omnu}{\omega_{\bar\nu}}
\newcommand{\ome}{\omega_{e}}
\newcommand{\sla}{\! \not \!}
\newcommand{\ep}{\epsilon}
\newcommand{\Real}{\Re{\rm e}}
\newcommand{\Img}{\Im{\rm m}}
\newcommand{\Gammamu}{\Gamma_{\mu}}
\newcommand{\Gammanu}{\Gamma_{\nu}}
\input epsf
\renewcommand{\topfraction}{0.8}


\title{
Direct Urca neutrino rate in colour superconducting quark matter}

\author{Prashanth Jaikumar} 
\affiliation{Physics Division, Argonne National Laboratory, 
             Argonne, IL 60439-4843 U.S.A.} 
             
\author{Craig D.\ Roberts} 
\affiliation{Physics Division, Argonne National Laboratory, 
             Argonne, IL 60439-4843 U.S.A.} 
\affiliation{Fachbereich Physik, Universit\"at Rostock, D-18051 Rostock, 
Germany} 

\author{Armen Sedrakian} 
\affiliation{Institute for Theoretical Physics, Universit\"at T\"ubingen, 
D-72076 T\"ubingen, Germany}

\begin{abstract}
If deconfined quark matter exists inside compact stars, the primary cooling mechanism is neutrino radiation via the direct Urca processes $d\to u + e+\anu_e$ and $u+e\to d+\nu_e$.  Below a critical temperature, $T_c$, quark matter forms a colour superconductor, one possible manifestation of which is a condensate of $\langle u d \rangle$ quark Cooper pairs in an electric-charge neutralising background of electrons.  We compute the neutrino emission rate from such a phase, including charged pair-breaking and recombination effects, and find that on a material temperature domain below $T_c$ the pairing-induced suppression of the neutrino emission rate is not uniformly exponential.  If gapless modes are present in the condensed phase, the emissivity at low temperatures is moderately enhanced above that of completely unpaired matter.  The importance of charged current pair-breaking processes for neutrino emission both in the fully gapped and partially gapped regimes is emphasised.
\end{abstract}


\maketitle


The direct Urca neutrino emission process (in-medium $\beta$-decay), which was originally introduced in studies of stellar collapse and supernova explosions \cite{GS}, plays a pivotal role in the cooling of compact stars \cite{Iwamoto1,Prakash1}.  When kinematically allowed it is one of the most efficient mechanisms by which massive neutron stars composed only of ordinary nuclear matter can cool, allowing them to reach temperatures below the observational threshold within roughly a decade \cite{fn1}.  

It is possible that the proton concentration in the interior of such a star may be insufficient to satisfy energy-momentum conservation in the decay $n\rightarrow p+e+\anu_e$.  However, if the interior contains \emph{unpaired} quark matter, then the reactions $d\to u + e+\anu_e$ and $u+e\to d+\nu_e$ can proceed easily since the quarks are ultrarelativistic, and interactions and/or finite quark masses guarantee that energy-momentum conservation is satisfied.  The reaction rate is proportional to the strong coupling constant $\alpha_s$ in the weak-coupling limit~\cite{Iwamoto1}.  These reactions provide for prodigious neutrino emission, which dominates over all other sources of neutrino production.  Hence, the presence of unpaired quark matter in a neutron star is expected to dramatically alter the star's cooling rate and consequently its temperature-age profile, in which case astronomical observations of this profile may be used to place constraints on the abundance of quark matter in stellar interiors.

This straightforward connection may be undermined if quarks condense to form Cooper pairs in the dense stellar interior.  (Aspects and consequences of such condensation are described in Refs.\ \cite{Wilczek}.) In this instance, one might anticipate that the direct Urca neutrino rate at temperatures $T<T_c$ will be greatly suppressed owing to the cost of bridging the pairing gap $\Delta(T)$ via thermal excitation.  ($T_c\lesssim 0.1\,$GeV is the temperature above which the pairing gap disappears.)
 
Herein we show that this is not necessarily the case because at moderate temperatures, $T\lesssim T_c$, modifications of the phase-space occupation factors resulting from the gap are non-exponential, and the breaking and formation of pairs also contributes to the neutrino rate.  In the following we describe a calculation that illustrates these features and supports a view that the suppression just below $T_c$ is not uniformly exponential.  This enables us to quantify the direct Urca neutrino rate in superconducting quark matter, which may be relevant to the cooling of compact stars.

The central densities of neutron stars may be as much as $(5-10) \rho_0$ depending on the underlying equation of state of dense matter; here  $\rho_0 = 0.16$ fm$^{-3}$ is the nuclear saturation density~\cite{wiringa}.  At such densities an ultrarelativistic two-flavour quark gas is characterised by a chemical potential $\mu_{u,d} \sim 0.4-0.5\,$GeV$\,\gtrsim \Lambda_{\rm QCD}$.  Perturbation theory is not a valid tool on this domain of $\mu_{u,d}$ and hence the true nature of the ground state of quark matter is thereupon uncertain.  Far from this scale, namely, at $\mu_{u,d}\gg \Lambda_{\rm QCD}$, perturbation theory predicts that quark matter exists in a colour-flavour-locked phase (CFL)~\cite{CFL}.  However, that is only one amongst many phases which may be realised as the chemical potential is reduced from asymptotic values to that characteristic of a neutron star interior \cite{PHASES}.  (NB.\ Measured in scales typical of QCD, neutron stars are cool; viz., $T \simeq 0$.)  

We will explore a minimal realistic scenario of two-flavour ($u$, $d$) quark matter, with spin-singlet channel pairing, constrained by charge neutrality and $\beta$-equilibrium.  Since increasing $\mu$ promotes quark-quark pairing, while increasing quark mass opposes it; viz., the magnitude of any putative gap, $\Delta(T)$, increases with increasing chemical potential but decreases with increasing mass, this picture can be valid when the relevant $s$-quark mass-scale \cite{fn2} satisfies $M_s^2(\mu,T) \gtrsim \mu \,\Delta(T)$.  For $\mu \sim 0.4\,$GeV and $\Delta(T) \lesssim 0.1\,$GeV~\cite{brs99} this means our scenario may be valid for $M_s (\mu,T) \gtrsim 0.2\,$GeV.  On this domain the $s$-quark is too heavy to influence pairing between the light-quarks.  Hence it plays no role in our subsequent analysis.  We note that $M_s(\mu\sim 0.4\,{\rm GeV},T<T_c) > 0.2\,$GeV is plausible~\cite{bastirev}.
 
It is plain that the enforcement of charge neutrality and $\beta$-equilibrium
entails a mismatch between the $u$- and $d$-quark chemical potentials;
i.e., $\delta\mu = (\mu_d-\mu_u)/2=\mu_e/2$.  The neutrino radiation rate therefore depends on the ratio $\zeta := \Delta/\delta\mu\,$.  On the domain $\zeta > 1$ the relevant quasiparticle dispersion law is gapped for all momentum modes.  However, for $\zeta<1$ there is a measurable domain of ungapped modes \cite{GAPLESS}.  Neutrino radiation in this regime has previously been studied for three flavour ($u,d,s$) pairing \cite{Jotwani}.  NB.\ A nonzero $\beta$-decay matrix element is only possible with a dressed-quark dispersion law.  Such modifications can arise through, e.g.: pairing; interaction effects that are linear in $\alpha_s$; a dynamically generated quark mass \cite{Iwamoto1}; and non-Fermi-liquid behaviour \cite{SS1}.  In the following we consider massless quarks, interaction corrections that are linear in $\alpha_s$, and the influence of pairing. 

Cooper pair breaking and formation processes play a central role during the neutrino radiation era both for superfluid neutron stars~\cite{SCHAAB} and compact stars featuring colour superconducting quark cores~\cite{Jotwani,BLASCHKE}.  The inclusion of pair-breaking and recombination effects below $T_c$ is therefore an important feature of our study.  The microscopic origin of the Cooper pair breaking and formation processes and their contribution to the neutrino emission from neutron star matter was explored in a number of works~\cite{Flowers,Vosky,Armen,Schwenk}. The counterparts of these processes in quark matter, i.e. the case where the neutrino emission is driven by weak neutral current interaction between quarks and neutrinos, was studied in Ref.~\cite{JP}. Herein we explore the role of the weak {\it charge current pair-breaking} process in neutrino radiation, which was found to be important in the nuclear Urca process~\cite{Armen}.

In order to calculate the total neutrino emission rate we note that, since the rates for $\beta$-decay and electron capture are identical in $\beta$-equilibrated matter, it is only necessary to compute one rate, say $d\rightarrow u+e+\anu_e$, and subsequently multiply the result by two.  We employ the Kadanoff-Baym formalism for neutrino transport \cite{knollvoskresensky,KB2}, wherein the neutrino emissivity, defined as the neutrino rate per unit volume, can be related to the $W$-boson polarisation tensor in the superconducting phase; namely
\begin{eqnarray}
\nonumber 
\epsilon_{\nu} & =&  -\tilde G_F^2 \int_{q_1,q_2}
\int\! \frac{d^4 \ell}{(2\pi)^4} \, \delta^4(\ell-q_1-q_2) \, \omega_2(\qvec_2)
  n_{B}(\ell_0)\\
&\times&  n_F(-\omega_1)n_F(-\omega_2)
 \Lambda^{\mu\lambda}(q_1,q_2) \, \Img\,\Pi^R_{\mu\lambda}(\ell)\,,
\label{nuemiss}
\end{eqnarray} 
where: $\tilde G = G_F\,\cos\, \theta_C$, with $G_F$ being the Fermi weak coupling constant and $\theta_C$ the Cabibbo angle ($\cos \theta_C = 0.973$);
$q_{1,2} = (\omega_{1,2},\bq_{1,2})$ are the 4-momenta of the leptons ($e$ and $\nu$, both considered massless and labelled, respectively, by indices 1 and 2) and 
$\ell$ is the $W$-boson momentum;
$\int_{q_i}:=\int {d^3q_i}/[(2\pi)^3\,2\,\omega_i]$ are phase-space integrals;
$n_B(\ell_0)$ is the Bose distribution function for the $W$-boson energy;
and $n_F(-\omega_{1,2})$ are the Pauli blocking factors for electrons and 
neutrinos in the final state.  (NB.\ Emission is characterised by $\omega_{1,2}>0$ and hence the Dirac delta function in Eq.~(\ref{nuemiss}) ensures that the integral does not have support for $\ell_0<0$). The lepton tensor in Eq.~(\ref{nuemiss}) is 
\begin{eqnarray} 
\nonumber
\lefteqn{\Lambda^{\mu\lambda}(q_1,q_2) = {\rm Tr}\left[\gamma^{\mu} 
(1 -\gamma^5)\sla q_1\gamma^{\lambda}(1-\gamma^5)\sla q_2\right]}\\
& = & 8 [q_1^{\mu} q_2^{\lambda} + q_2^{\mu} q_1^{\lambda} - q_1\cdot q_2\, 
g^{\mu\lambda} + i\epsilon^{\mu\alpha\lambda\beta} \,q_{1\alpha}
q_{2\beta}]\,. 
\label{leptensor}
\end{eqnarray}

\begin{figure}[t]
\begin{center}
\includegraphics[width=0.45\textwidth]{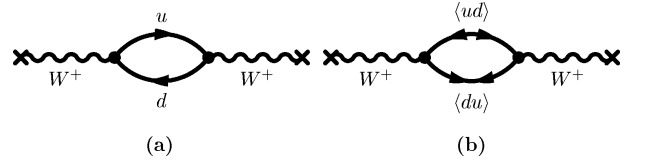}
\end{center}
\vspace*{-4ex}

\caption{One-loop $W$-polarisation tensor: (a) normal piece, which is the sole contribution above $T_c$; and (b) ``anomalous'' piece, which is proportional to $\Delta^2(T)$ and vanishes at $T_c$.  The imaginary part of (b) (nonzero below $T_c$) contributes the $\nu$-rate from charged-pair breaking and formation.}
\label{fig1}
\end{figure}

At one-loop level, the retarded $W$-boson polarisation tensor in the superconducting phase, $\Pi^R_{\mu\lambda}(k)$, is obtained by combining the two physically distinct contributions illustrated in Fig.\ \ref{fig1}.  To evaluate this tensor, we require expressions for the quark propagators in the two-flavour superconductor.  These are detailed in \cite{Shov1}, where the gluon polarisation was studied.  For our purpose, we replace the gluon vertex by the weak vertex $\Gamma_{\pm}^{\mu} = \gamma^{\mu} (1-\gamma_5) \otimes \tau_{\pm}$, where $\tau_{\pm}=(\tau_1\pm\tau_2)/2$ are flavour-raising and lowering operators constructed from Pauli matrices.  In addition, antiparticle contributions are neglected because we work at large quark chemical potential.  
 
It follows that diagram (a) in Fig.\ \ref{fig1} represents 
\begin{eqnarray}
\nonumber
&&\Pi^{R,(a)}_{\mu\lambda}(\ell) = -2T \int \frac{d^3 {\bf p}}{(2\pi)^3}
\frac{d^3 {\bf k}}{(2\pi)^3}\,
\delta^3({\bf  k-p-\mbox{\boldmath $\ell$}})\, \\ 
&& \hspace{-1cm} \times  \sum_{n} \,H^{(a)}_{\mu\lambda} 
\frac{(k_{0n}+\delta\mu+E_k^-)(p_{0n}-\delta\mu+E_p^-)}
{[(k_{0n}+\delta\mu)^2-{\xi_k^-}^2][(p_{0n}-\delta\mu)^2-{\xi_p^-}^2]} %
 \,, \label{Piline1}
\end{eqnarray}
with 
\begin{equation}
H^{(a)}_{\mu\lambda} = {\rm Tr}[\gamma_{\mu}(1-\gamma_5)\Lambda^+({\bf p})
\gamma_0\gamma_{\lambda}(1-\gamma_5)\Lambda^+({\bf k})\gamma_0]\,, 
\label{hatensor}
\end{equation}
where the trace is over colour, flavour and spinor indices and 
$
\Lambda^{\pm}({\bf k})= ( 1 \pm {\bf\gamma}\cdot\hat{\bf k})/2.
$ 
In Eq.~(\ref{Piline1}), $E_k^-=k-\bar{\mu}$ and $E_p^-=p-\bar{\mu}$, with $\bar{\mu}=(\mu_d+\mu_u)/2$, $\xi_{k,p}^-=\sqrt{{E_{k,p}^-}^2+\Delta^2(T)}$, and the Matsubara sum is over discrete frequencies $p_{0n}=(2n+1)i\pi T$, with $k_{0n}=p_{0n}+\ell_0$.  We observe that in the normal phase, defined by $\Delta\equiv 0$, 
$
\Lambda^{\mu\lambda}H^{(a)}_{\mu\lambda} = 64 \, \omega_1 \omega_2\, (1-\hat{\bf q}_1\cdot\hat{\bf k})(1-\hat{\bf q}_2\cdot\hat{\bf p}) \equiv 0\,,
$
\emph{unless} the quark masses are nonzero or one includes perturbative corrections to the free-quark dispersion relation.  This follows from four-momentum conservation, which for massless free-quarks requires all four particles to be collinear; viz., $\hat{\bf q}_1\cdot\hat{\bf k} = 1 = \hat{\bf q}_2\cdot\hat{\bf p}$.  Once interactions modify the quark dispersion relation, this is no longer true, and \cite{Iwamoto1} 
\begin{equation}
\Lambda^{\mu\lambda}H^{(a)}_{\mu\lambda} = 64 \, \omega_1 \omega_2\, \frac{16 \alpha_s}{3 \pi} 
\label{collinearalpha}
\end{equation}
provides the leading-order perturbative contribution from diagram (a) to $\epsilon_\nu$ in Eq.\,(\ref{nuemiss}).

Turning to the anomalous contribution, (b) in Fig.\ \ref{fig1}, which has gap insertions in the quark lines, we obtain
\begin{eqnarray}
\nonumber
\lefteqn{ \Pi^{R,(b)}_{\mu\lambda}(\ell) = 2T\Delta^2(T) \int \frac{d^3 {\bf p}}{(2\pi)^3}
\frac{d^3 {\bf k}}{(2\pi)^3}\,
\delta^3({\bf  k-p-\mbox{\boldmath $\ell$}})\, }\\ 
&& \hspace{-1cm} \times\,  \sum_{n}\,H^{(b)}_{\mu\lambda}\, \frac{1}
{[(k_{0n}+\delta\mu)^2-{\xi_k^+}^2][(p_{0n}+\delta\mu)^2-{\xi_p^-}^2]},
\label{anomalous}
\end{eqnarray}
with
\begin{equation}
H^{(b)}_{\mu\lambda}={\rm Tr}[\gamma_{\mu}(1-\gamma_5)\Lambda^+({\bf p})\gamma_5\gamma_{\lambda}(1+\gamma_5)\Lambda^+({\bf k})\gamma_5], \label{hbtensor}
\end{equation}
where $\xi_{k}^+=\sqrt{{E_k^+}^2+\Delta^2(T)}$, with $E_k^+=k+\bar{\mu}$.  In Eqs.~(\ref{Piline1}) and (\ref{anomalous}), the trace over the colour-flavour structure of the condensate yields a factor of two.  
 
The material features of the anomalous contribution, Eq.~(\ref{anomalous}), are that it appears only in the superconducting state and vanishes in the limit $T\rightarrow T_c^-$.  For our argument and illustration it is sufficient to employ a parameterization of the gap~\cite{Lindblom} that is based on the BCS model calculation of Ref.\,\cite{Muhlschlegel}:
\begin{equation}
\Delta(T) = \Delta(0)\left[{1 - \left({T}/{T_c}\right)^{3.4}}\right]^{0.53}
\theta(T_c-T)\,.
\end{equation}
While modifications to this profile have a modest quantitative impact on our results, they are immaterial to our primary conclusions.
 
Since we have simple algebraic expressions for each element that appears in Eqs.~(\ref{Piline1}) and (\ref{anomalous}), it is straightforward to evaluate the Matsubara sums and subsequently to obtain the imaginary part of the retarded polarisation tensor via analytic continuation \cite{fn3}; i.e., $\ell_0\rightarrow \ell_0 + i\epsilon$. From Eq.~(\ref{Piline1}) one obtains the normal contribution:
\begin{eqnarray}
&&\hspace{-0.7cm} \Img\,\Pi^{R,(a)}_{\mu\lambda}(\ell)  
= -\frac{\pi}{2} \int\! \frac{d^3 {\bf p}}{(2\pi)^3} \, 
H^{(a)}_{\mu\lambda}\nonumber\\
&\times& 
\bigl[(f_p^{-,+} - f_{p+\ell}^{-,-})\, 
\delta_n(\ell_{0}^{+\,-})\, 
({\cal E}_{p+\ell}^- + 1)\, ({\cal E}_{p}^- + 1)\nonumber\\
&-&(1-f_p^{-,-} -f_{p+\ell}^{-,-})\, \delta_n(\ell_{0}^{-\,-})\,
({\cal E}_{p+\ell}^- +1 ) \, ({\cal E}_{p}^- -1 )\nonumber\\
&-& (f_p^{-,-} - f_{p+\ell}^{-,+}) 
\,\delta_n(\ell_{0}^{-\,+}) \,({\cal E}_{p+\ell}^- -1 ),\, 
({\cal E}_{p}^- - 1)\bigr],
\label{Impia}
\end{eqnarray}
where $f_{r}^{\pm,\alpha} = f(\xi_r^{\pm}+\alpha\delta\mu)$, $r =  p,p+\ell$, are equilibrium free particle Fermi distribution functions, $\delta_n(\ell_{0}^{\alpha\beta}) = \delta(\ell_0+\alpha\xi_p^-+ \beta\xi_{p+\ell}^-+2\delta\mu)$ and ${\cal E}_r^{\pm} = E_r^{\pm}/\xi_r^{\pm}$.  The anomalous contribution is obtained from Eq.~(\ref{anomalous})
\begin{eqnarray}
&&\hspace{-0.7cm}
\Img\,\Pi^{R,(b)}_{\mu\lambda}(\ell)  = -\frac{\pi}{2}\Delta^2
\int\! \frac{d^3 {\bf p}}{(2\pi)^3}\, 
H^{(b)}_{\mu\lambda}\frac{1}{\xi_p^-\, \xi_{p+\ell}^+} \nonumber\\
&\times& 
\bigl[(f_p^{-,-}-f_{p+\ell}^{+,-}) \, \delta_a(\ell_{0}^{+\,-}) 
- ( f_{p+\ell}^{+,-} - f_p^{-,+}) \, \delta_a(\ell_{0}^{-\,+}) \nonumber\\
&-& (1 - f_{p+\ell}^{+,-} - f_p^{-,+})
\, \delta_a(\ell_{0}^{-\,-})\bigr],
\label{Impib}
\end{eqnarray}
where $\delta_a(\ell_0^{\alpha\beta}) = \delta(  \ell_0 + \alpha \xi_p^- + \beta\xi_{p+\ell}^+)$. 

The Dirac-delta functions in Eqs.~(\ref{Impia}) and (\ref{Impib}) make the kinematics of scattering and pair breaking transparent.  The $\delta(l_0^{+\,-})$ and $\delta(l_0^{-\,+})$ terms describe scattering of $u$- and $d$-quark quasiparticles from the condensate.  The $\delta(l_{0}^{-\,-})$ terms, on the other hand, express breaking of charged Cooper pairs when the total energy $\ell_0$ exceeds the pair threshold; viz., $\ell_0 > 2\Delta$.  The term proportional to $\delta(l_0^{+\,+})$ has been dropped as it does not contribute to the emission rate. The delta functions also simplify the evaluation of the integrals in Eqs.\,(\ref{Impia}) and (\ref{Impib}).  More importantly, they place bounds on the magnitude of $\mbox{\boldmath $p$}^2$ so that the integrals are finite.
 
We now return to the neutrino emission rate of Eq.\,(\ref{nuemiss}).  In addition to Eq.~(\ref{collinearalpha}) and at the same order we have
\begin{eqnarray}
\nonumber 
\lefteqn{\Lambda^{\mu\lambda}H^{(b)}_{\mu\lambda} = 32\omega_1\omega_2 \bigg[ (1 + \hat {\bf p} \cdot \hat {\bf k} ) 
- \frac{16 \alpha_s}{3 \pi}\,\hat {\bf p}\cdot \hat {\bf q}_2 
}\\
&- & 
 \hat {\bf q}_1 \hat {\bf q}_2 (1 - \hat {\bf p} \cdot \hat {\bf k} )
- \hat {\bf p}\cdot \hat {\bf q}_1\, (1 + \hat {\bf k} \cdot \hat {\bf q}_2 ) 
\bigg]
%
%
\end{eqnarray}
Thus
\begin{eqnarray}
\nonumber
\epsilon_{\nu} &=& -\, \tilde G_F^2 \int_{q_1,q_2}
\int\! \frac{d^4 \ell}{(2\pi^4)} \, \delta^4(\ell-q_1-q_2) \,
 \omega_2(\qvec_2)\\ 
&\times&   \, 
 n_{B}(\ell_0) \, n_F(-\omega_1)n_F(-\omega_2) \, {\cal F}(\ell_0,\mbox{\boldmath $\ell$},q_1,q_2)\,,
\label{nuemiss2}
\end{eqnarray}
with 
\begin{eqnarray}
\lefteqn{{\cal F}(\ell_0,\mbox{\boldmath $\ell$},q_1,q_2)
=\Lambda^{\mu\lambda}(q_1,q_2) \, \Img\,\Pi^R_{\mu\lambda}
(\ell_0,\mbox{\boldmath $\ell$})}\nonumber\\
&=&\Lambda^{\mu\lambda}(q_1,q_2)\left[ 
\Img\,\Pi^{R,(a)}_{\mu\lambda}(\ell)+
\Img\,\Pi^{R,(b)}_{\mu\lambda}(\ell)\right]\!.
\end{eqnarray}
 
It is straightforward to evaluate the neutrino emission rate in Eq.~(\ref{nuemiss2}).  One of the integrations, say that over the neutrino's three-momentum, is made trivial by the factor $\delta^3(\ell-q_1-q_2)$:  $\mbox{\boldmath $q$}_2 = \mbox{\boldmath $\ell$} - \mbox{\boldmath $q$}_1$, which  leaves $\delta(\ell_0 - \omega_1 -\omega_2)$ with $\omega_2 = |\mbox{\boldmath $q$}_2| = |\mbox{\boldmath $\ell$} - \mbox{\boldmath $q$}_1|$.  This last distribution bounds the accessible domain of $|\mbox{\boldmath $q$}_1|$; namely, 
$\ell_0 - |\mbox{\boldmath $\ell$}| < 2 |\mbox{\boldmath $q$}_1|
< \ell_0 + |\mbox{\boldmath $\ell$}|$, 
and fixes the value of  
$\mbox{\boldmath $\ell$} \cdot \mbox{\boldmath $q$}_1$ 
within this domain.  It remains to evaluate the integrals over $|\mbox{\boldmath $q$}_1|$, $\mbox{\boldmath $\ell$}$ and $\ell_0$, tasks we perform numerically, in the order listed.  NB.\ The Bose factor $n_B(\ell_0)$ guarantees rapid convergence of the $\ell_0$ integral.

\begin{figure}[t]
\begin{center}
\includegraphics[width=0.45\textwidth]{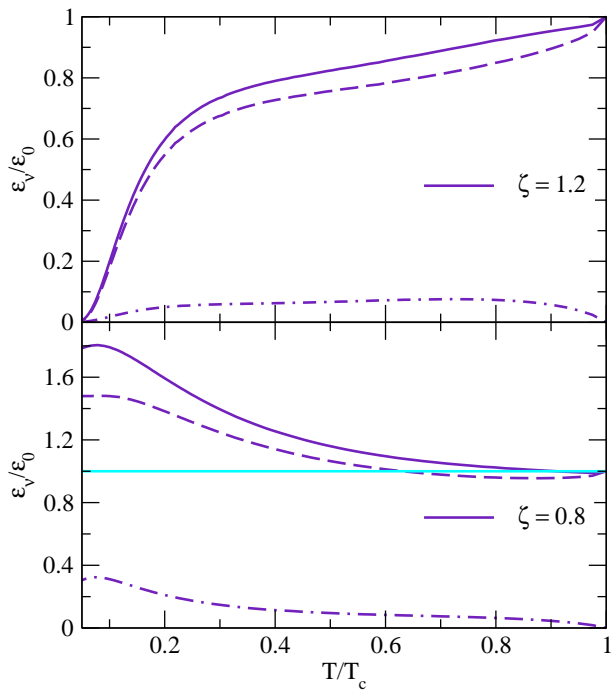}
\end{center}
\caption{Temperature dependence of the emissivity Eq. (\ref{nuemiss2}) normalized to its value at the critical temperature.  Upper panel -- fully gapped regime, $\zeta>1$; and lower panel -- partially gapped regime, $\zeta <1$.  In both panels the \textit{dashed} and \textit{dash-dotted} curves are, respectively, the normal and anomalous contributions; and the \textit{solid} line is their sum.  In the lower panel, for reference, the horizontal line $\epsilon_{\nu}/\epsilon_0=1$ is the result for unpaired matter.
The curves were obtained with $\mu_d = 400$, $\mu_u = 300$ and $\mu_e = \mu_d -\mu_u = 100\,$MeV, values which are typical of two-flavour colour superconducting quark matter.}
\label{fig2}
\end{figure}

Our calculated result for the emissivity of superfluid quark matter at $T=T_c$ matches that of the unpaired state \cite{Iwamoto1}: 
\begin{equation}
\label{iwamotoeq}
\epsilon_{0}= ({914}/{315})~\tilde G_F^2~\alpha_s~\mu_d~\mu_u~\mu_e~T^6\,,
\end{equation}
to within a few percent.  The small deviation owes to our more complete treatment of nonzero temperature effects.  

In Fig.\,\ref{fig2} we depict the ratio $\epsilon_{\nu}/ \epsilon_{0}$ as a function of $t= T/T_c$.  The upper panel makes plain that neutrino production is suppressed in the presence of complete pairing; i.e., when all momentum modes are gapped.  However, when the magnitude of the gap is moderate, that suppression is not uniformly exponential.  Rather, it is approximately linear on a material domain.  For the parameters we have chosen, that domain is $0.4\leq T/T_c \leq 1$.  Naturally, at any given value of $t$ the emissivity decreases with increasing $\zeta$: a larger gap must be overcome.
In the fully-gapped regime, for straightforward kinematic reasons, the anomalous contribution vanishes in the limits $t\to 0$ and $t\to 1$.

The emissivity in the partially-gapped regime is illustrated in the lower panel of Fig.\,\ref{fig2}: $\forall\, \zeta \in (0,1)$ the neutrino emission is enhanced cf.\ Eq.\,(\ref{iwamotoeq}).  The normal contribution is magnified because the gap persists for a measurable set of momentum modes and hence more phase space is accessible to these modes than would otherwise be the case.  The effect disappears as $\zeta \to 0$. The behaviour of the anomalous contribution in this regime is also understandable: it is nonzero but decreases uniformly in magnitude as $\zeta\to 0$, driven by the reduction in phase space and the overall multiplicative factor of $\Delta^2$ in Eq.\,(\ref{anomalous}).  The result for unpaired matter is recovered at $t=0$.  

In the neighbourhood of $t=0$
\begin{equation}
\epsilon_{\nu}/\epsilon_0 \propto
\{C(\zeta)+{\rm exp}[(\zeta-1)\delta\mu/T]\}^{-1},
\end{equation}
where $C(\zeta)\sim {\rm O}(1)$ is independent of temperature; viz., in the fully-gapped regime the normalised emissivity is exponentially suppressed, while in the partially-gapped regime it tends to a constant.

We have considered isospin singlet pairing and the behaviour of $\epsilon_\nu$ exhibited herein differs from that found in earlier studies of isospin-triplet pairing; viz., Refs.\,\cite{DGY,Yakovlev}, which focused on reaction phase space, omitted the coherence factors and pair-breaking processes, and used a non-BCS dispersion relation; and Ref.\,\cite{Armen}, which included those effects along with a BCS dispersion relation.  

Our calculation in the two-flavour phase is indicative of similar behaviour in the neutrino rate below $T_c$ for the three-flavour superconductor (CFL phase).  However, in that case quark quasiparticle excitations play a role secondary to CFL Goldstone bosons so far as neutrino emission and cooling of stars with a CFL core are concerned
\cite{JPS,RST}.  As a follow up to the present study, it may be useful to determine the neutrino opacity in a two-component quark superfluid.  This, determined by scattering and absorption of neutrinos, might be accomplished by following the pattern of Ref.~\cite{Kundu}.  Neutrino emission via the Urca process from a spin-one colour superconductor has also been assessed \cite{Schmitt,Schmitt:2005wg}.  Neutral current mediated pair-breaking processes would be relevant to spin-one phases of paired quark matter.

We focused on charged-pair breaking because the neutrino emission rate generated by direct Urca is far greater than that provided by the neutral analogue (neutrino bremsstrahlung).  In addition, for the two-flavour case there are no charged Goldstone modes that can couple to neutrinos.  The contribution to neutral-current neutrino scattering and emission from such soft ($\omega \ll \Delta$) collective excitations was highlighted for the CFL phase in Ref.~\cite{Kundu}.  For example, their inclusion is essential to an internally consistent truncation of the Dyson-Schwinger equations and
the concomitant preservation of relevant Ward-Takahashi identities.  As noted, such contributions are absent in the discussion of a two-flavour superconductor and one 
can unambiguously obtain the transverse part of the $W$-boson polarisation tensor without model-dependent assumptions about the dressed weak vertex.  

To close, we note that the temperatures within a star during the protoneutron phase of its evolution may support a colour superconducting state in its dense core because the critical temperature is anticipated to be large: $T_c \approx 0.6 \Delta_0 \sim 50\,$MeV.  Neutrinos would interact often inside quark matter and therefore serve as an electroweak probe of the superconducting phase.  Numerical simulations of the early stages in the evolution of newly-born hot neutron stars indicate that these systems support a lepton fraction of $\approx 0.4$ \cite{proto}.  With the neutrinos trapped, this fraction represents electrons and neutrinos.  

Now, should a transition to neutral quark matter take place at this early stage, the system can satisfy the condition of $\beta$-equilibrium ($\mu_d - \mu_u = \mu_e - \mu_{\nu_e}$) while supporting an electron fraction $x_e\approx 0.2$.  (NB.\ Neutrinos free stream in a cool star, in which case $\mu_{\nu_e}\approx 0$ and $x_e\approx 0.01$.)  Furthermore, recent model estimates of the equation of state for $\beta$-equilibrated charge-neutral matter suggest that in the dense core of a neutron star, despite its positive charge, two- or three-flavour Cooper-paired quark matter can occupy a large fraction of the volume as one component of a mixed phase that also contains normal quark matter \cite{Shovkovy2}.  

Considered together with neutrino opacities, which are large at $T\sim T_c$, it is apparent that the results illustrated by Fig.~\ref{fig2} are pertinent to any detailed computation of the cooling history of a star that undergoes a phase transition from nuclear matter to superconducting quark matter.

%
We acknowledge interactions with M.\,G.~Alford, D.\,B.~Blaschke, A.~H\"oll, A.~Schmitt, I.~Shovkovy, Q.~Wang and S.~V.~Wright; and thank Helmholtz Association Virtual Institute VH-V1041 ``Dense Hadronic Matter and QCD Phase Transitions'' for supporting the meeting at which this study was conceived.
This work was supported by: Department of Energy, Office of Nuclear Physics, contract no.\ W-31-109-ENG-38; National Science Foundation contract no.\ INT-0129236; Deutsche 
Forschungsgemeinschaft SFB 382; and the \textit{A.\,v.\ Humboldt-Stiftung} 
via a \textit{F.\,W.\ Bessel Forschungspreis}.
\vspace*{\fill}
%



\begin{thebibliography}{00} 
\bibitem{GS}
G.\ Gamow and M.\ Schoenberg, Phys.\ Rev.\ {\bf 59},  539 (1941).

\bibitem{Iwamoto1}
N.~Iwamoto,
Phys.\ Rev.\ Lett.\  {\bf 44}, 1637 (1980);
A.~Burrows,
\emph{ibid}.\ 1640.

\bibitem{Prakash1} 
J.\,M.~Lattimer, M.~Prakash, C.\,J.~Pethick and P.~Haensel,
Phys.\ Rev.\ Lett.\  {\bf 66}, 2701 (1991);
M.~Prakash, M.~Prakash, C.\,J.~Pethick and J.\,M.~Lattimer, ApJ {\bf 390}, L77 (1992).

\bibitem{fn1} This and other aspects of the physics of neutron stars are described in ``Physics of Neutron Star Interiors,'' eds.\ D.~Blaschke, N.\,K.\ Glendenning and A.\ Sedrakian, Lecture Notes in Physics \textbf{578} (Springer-Verlag, New York, 2001).
 
\bibitem{Wilczek} K.~Rajagopal and F.~Wilczek,
hep-ph/0011333; 
M.~Alford,
Prog.\ Theor.\ Phys.\ Suppl.\  {\bf 153}, 1 (2004);
D.\,H.~Rischke,
Prog.\ Part.\ Nucl.\ Phys.\  {\bf 52}, 197 (2004)

\bibitem{wiringa} R.~B.~Wiringa, V.~Fiks and A.~Fabrocini,
Phys.\ Rev.\ C {\bf 38}, 1010 (1988);
N. K. Glendenning, {\it Compact Stars}, 
2nd ed.\ (Springer-Verlag, New York, 2000);
%
F. Weber, {\it Pulsars as Astrophysical Laboratories for Nuclear and
Particle Physics}, \ (IOP Publishing, Bristol, 1999).

\bibitem{CFL} M.\,G.~Alford, K.~Rajagopal and F.~Wilczek,
Nucl.\ Phys.\ B {\bf 537}, 443 (1999).
  
\bibitem{PHASES} M.\,G.~Alford, J.\,A.~Bowers and K.~Rajagopal,
Phys.\ Rev.\ D {\bf 63}, 074016 (2001); 
H.~Muther and A.~Sedrakian,
Phys.\ Rev.\ D {\bf 67}, 085024 (2003);
%
R.~Casalbuoni and G.~Nardulli, Rev. Mod. Phys. {\bf 76}, 263 (2004);  
M.\,G.~Alford, C.~Kouvaris and K.~Rajagopal,
Phys.\ Rev.\ Lett.\  {\bf 92}, 222001 (2004).

\bibitem{fn2} The relevant scale is provided by the $s$-quark
mass function.  In QCD the quark mass function is momentum dependent
and also depends on chemical potential and temperature.  These
features are discussed in Ref. \cite{bastirev}.            

\bibitem{brs99} J.\,C.\,R.~Bloch, C.\,D.~Roberts and S.\,M.~Schmidt,
Phys.\ Rev.\ C {\bf 60}, 065208 (1999).

 \bibitem{bastirev} C.~D.~Roberts and S.~M.~Schmidt,
Prog.\ Part.\ Nucl.\ Phys.\  {\bf 45}, S1 (2000); and references therein. 

\bibitem{GAPLESS} I.~Shovkovy and M.~Huang,
Phys.\ Lett.\ B {\bf 564}, 205 (2003).
  
\bibitem{Jotwani} M.~Alford, P.~Jotwani, C.~Kouvaris, J.~Kundu and 
K.~Rajagopal,
Phys.\ Rev.\ D {\bf 71}, 114011 (2005).
  
\bibitem{SS1} T.~Sch\"afer and K.~Schwenzer,
Phys.\ Rev.\ D {\bf 70}, 114037 (2004);
\textit{ibid.}, 054007 (2004).

\bibitem{SCHAAB} 
C.~Schaab, D.~Voskresensky, A.\,D.~Sedrakian, F.~Weber and M.\,K.~Weigel,
Astron.\ Astrophys.\  {\bf 321}, 591 (1997);
D.~Page, J.\,M.~Lattimer, M.~Prakash and A.\,W.~Steiner, Ap.\ J.\ Suppl. \textbf{155}, 623 (2004); 
%
D.~Blaschke, H.~Grigorian and D.\,N.~Voskresensky,
Astron.\ Astrophys.\  {\bf 424}, 979 (2004).

\bibitem{BLASCHKE} H.~Grigorian, D.~Blaschke,  and D.\,N.~Voskresensky,
Phys.\ Rev.\ C {\bf 71}, 045801 (2005).

\bibitem{Flowers} E.~Flowers, M.~Ruderman and P.~Sutherland, 
Astrophys.\ J.\ {\bf 205}, 541 (1976).

\bibitem{Vosky} D.\,N.~Voskresensky and A.\,V.~Senatorov,
  Sov.\ J.\ Nucl.\ Phys.\  {\bf 45}, 411 (1987).
  [Yad.\ Fiz.\  {\bf 45}, 657 (1987)].

\bibitem{Armen} A.~Sedrakian,
Phys.\ Lett.\ B {\bf 607}, 27 (2005).

\bibitem{Schwenk} A.~Schwenk, P.~Jaikumar and C.~Gale,
Phys.\ Lett.\ B {\bf 584}, 241 (2004).

\bibitem{JP} P.~Jaikumar and M.~Prakash,
Phys.\ Lett.\ B {\bf 516}, 345 (2001).

\bibitem{knollvoskresensky} J.~Knoll and D.\,N.~Voskresensky,
Annals Phys.\  {\bf 249}, 532 (1996).

\bibitem{KB2} A.~Sedrakian and A.~Dieperink,
Phys.\ Lett.\ B {\bf 463}, 145 (1999);
Phys.\ Rev.\ D {\bf 62}, 083002 (2000).

\bibitem{Shov1} M.~Huang and I.~A.~Shovkovy,
Phys.\ Rev.\ D {\bf 70}, 094030 (2004).

\bibitem{Lindblom} L.~Lindblom and B.~J.~Owen,
Phys.\ Rev.\ D {\bf 65}, 063006 (2002).

\bibitem{Muhlschlegel} B. M\"uhlschlegel, Z. Physik, {\bf 155}, 313 (1959).

\bibitem{fn3} In general the propagator for a dressed-quark differs signficantly from that of a free fermion; e.g., even in the
absence of dynamical chiral symmetry breaking, the vector piece of the
quark self-energy is nonzero \cite{brs99}.  In this case techniques
must be developed that facilitate a numerical evaluation of the
Matsubara sum; e.g., Refs.\ \cite{NumMatsubara}.
 
\bibitem{NumMatsubara} D.~Blaschke, C.\,D.~Roberts and S.\,M.~Schmidt,
Phys.\ Lett.\ B {\bf 425}, 232 (1998);
D.~Gomez Dumm and N.~N.~Scoccola,
Phys.\ Rev.\ C {\bf 72}, 014909 (2005).

\bibitem{DGY} D.\,G.~Yakovlev, A.\,D.~Kaminker, O.\,Y.~Gnedin and P.~Haensel,
Phys.\ Rept.\  {\bf 354}, 1 (2001).

\bibitem{Yakovlev} K.\,P.~Levenfish and D.\,G.~Yakovlev, Astron.\ Lett.\ {\bf 20}, 43 (1994); 
%
D.\,G.~Yakovlev, K.\,P.~Levenfish and Y.\,A.~Shibanov,
Phys.\ Usp.\  {\bf 169}, 825 (1999)
  
\bibitem{JPS} P.~Jaikumar, M.~Prakash and T.~Sch\"afer,
Phys.\ Rev.\ D {\bf 66}, 063003 (2002).

\bibitem{RST} S.~Reddy, M.~Sadzikowski and M.~Tachibana,
Nucl.\ Phys.\ A {\bf 714}, 337 (2003).

\bibitem{Kundu} J.~Kundu and S.~Reddy,
Phys.\ Rev.\ C {\bf 70}, 055803 (2004).

\bibitem{Schmitt} A.~Schmitt,
Phys.\ Rev.\ D {\bf 71}, 054016 (2005).

\bibitem{Schmitt:2005wg}
  A.~Schmitt, I.~A.~Shovkovy and Q.~Wang,
  arXiv:hep-ph/0510347.

\bibitem{proto} A.~Burrows and J.\,M.~Lattimer,
Astrophys.\ J.\  {\bf 307}, 178 (1986);
W.~Keil and H.-T.~Janka,
Astron.\ Astrophys.\  {\bf 296}, 145 (1995).

\bibitem{Shovkovy2} S.\,B.~Ruster, V.~Werth, M.~Buballa, I.\,A.~Shovkovy and D.\,H.~Rischke,
Phys.\ Rev.\ D {\bf 72}  034004 (2005);
D.~Blaschke, S.~Fredriksson, H.~Grigorian, A.\,M.~Oztas and F.~Sandin,
Phys.\ Rev.\ D {\bf 72} 085024  (2005).
  
\end{thebibliography}
\end{document}